\documentclass[fleqn,10pt]{wlscirep} 
\usepackage{graphicx}
\pdfoutput=1
\usepackage{textcomp}
\usepackage{subfig}

\title{Charge filling factors in clean and disordered arrays of tunnel junctions}

\author[1,*]{Kelly A. Walker}
\author[1]{Nicolas Vogt}
\author[1,$\dag$]{Jared H. Cole}
\affil[1]{Chemical and Quantum Physics, School of Applied Sciences, RMIT University, Melbourne, 3001, Australia}

\affil[*]{kelly.walker@rmit.edu.au}
\affil[$\dag$]{jared.cole@rmit.edu.au}

\begin{abstract}
We simulate one-dimensional arrays of tunnel junctions using the kinetic Monte Carlo method to study charge filling behaviour in the large charging energy limit. By applying a small fixed voltage bias and varying the offset voltage, we investigate this behaviour in clean and disordered arrays (both weak and strong disorder effects). The offset voltage dependent modulation of the current is highly sensitive to background charge disorder and exhibits substantial variation depending on the strength of the disorder. We show that while small fractional charge filling factors are likely to be washed out in experimental devices due to strong background charge disorder, larger factors may be observable.
\end{abstract}
\begin{document}

\flushbottom
\maketitle
\thispagestyle{empty}

\section*{Introduction}
Since the prediction\cite{BenJacob:1985, Averin:1986,Likharev:1989} and subsequent observation\cite{Delsing:1989ea,Delsing:1990ik} of single charge tunnelling in small metallic tunnel junctions, the field of low temperature nanoelectronics has developed rapidly. Fabrication of ultrasmall metallic islands allows the discrete nature of the island charge states to be observed. This leads to the suppression of conduction at low bias voltages, due to Coulomb blockade. However, introducing multiple junctions increases the complexity of the charge dynamics significantly. Exploring and understanding the structure and dynamics of low energy excitations in multi-junction systems and the effect of random offset charges has become important in improving the accuracy and design of these circuits.

\begin{figure}[th!]  
\centering 
\includegraphics[width=0.8\textwidth]{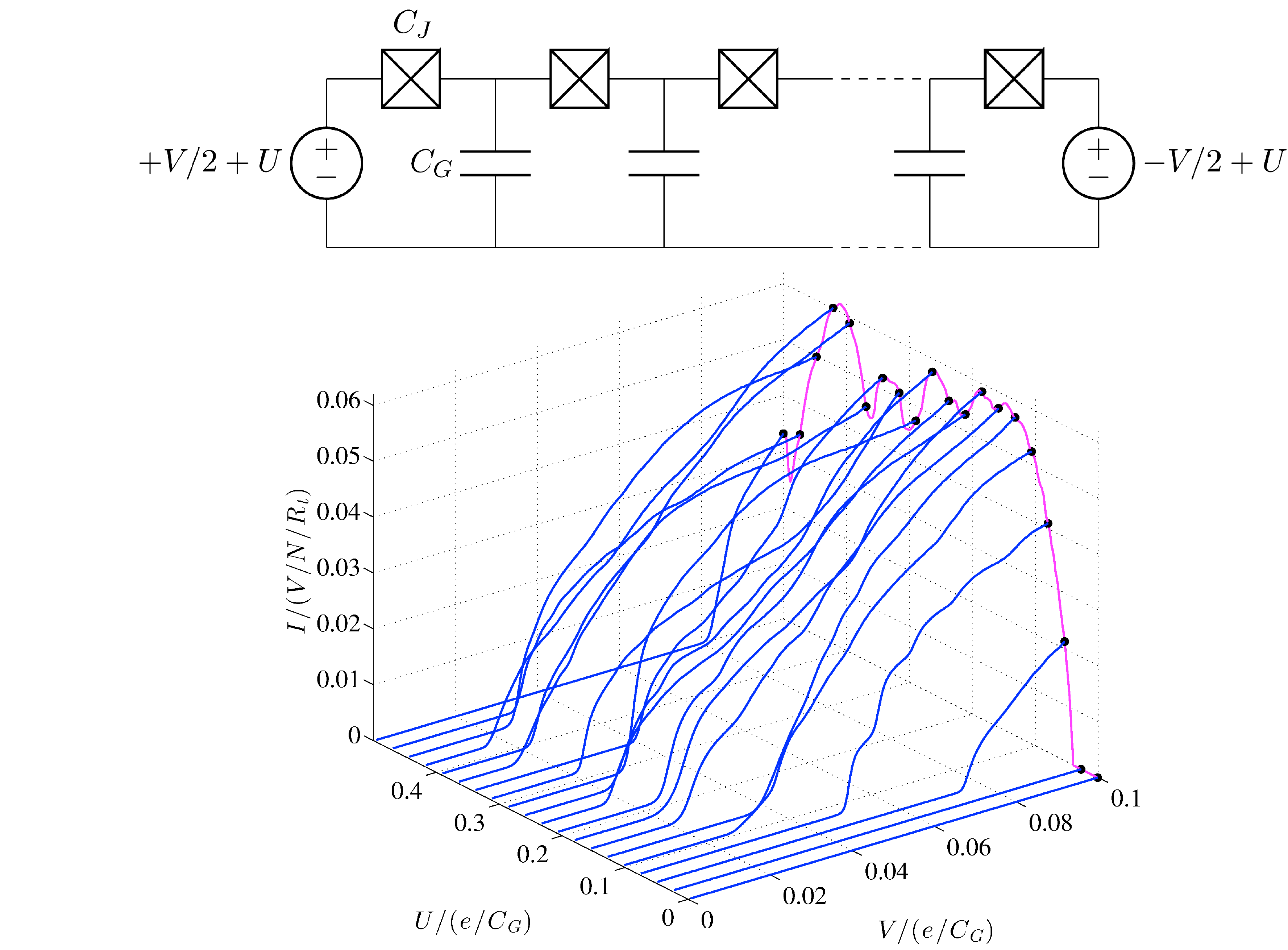}
\caption{\label{fig:3d_IVU_withdots}Top: Schematic of the junction circuit we simulate, consisting of a linear chain of $N+1$ tunnel junctions each with junction capacitance $C_J$. A symmetric bias $V$ is applied across the array with an additional offset $U$ relative to ground and each island is capacitively coupled to a ground plane through $C_G$. Bottom: $I$-$V$-$U$ characteristics for 20 different offset voltages from $U=0$ to $U=0.5$ $e/C_G$. Due to the voltage offset $U$, the $I$-$V$ responses are not simply linear within this $V$ range and the threshold voltage is also $U$-dependent. The current at fixed offset voltage bias is also plotted (magenta line), for $V=0.1$ $e/C_G$.
}
\end{figure}

An example of such a multi-junction circuit is a tunnel junction array which consists of a chain of small metallic islands separated by tunnel barriers, see Fig.~\ref{fig:3d_IVU_withdots}. Transport through the array can be correlated due to the interplay between the bias voltage across the array and the Coulomb repulsive interaction between charges. This interaction decays exponentially over a characteristic length $\Lambda$ that is determined by the ratio of the circuit capacitances. Consequently, in the low bias regime, charges within a junction array tend to be equidistant and in long arrays, at the onset of charge injection, a charge pattern quickly emerges. 

These patterns are dependent on the charge filling within the array, which is an additional modifier of the current. Different patterns emerge when certain filling factors are enforced by applying an offset voltage $U$, which changes the chemical potential, resulting in variation in the charge periodicity within the array. For example, a specific charge configuration within the array can mean the difference between a maximal current signal and current quenching.

Owing to their nanoscale geometry and as a result, large charging energy, these systems are highly sensitive to even small fractional random offset charges in the surrounding substrate. The effects of random charge disorder have been investigated in various single charge devices including transistors\cite{Starmark:1999,Krupenin:2000} and arrays\cite{Middleton:1993,Melsen:1997,Matsuoka:1998,Mueller:1998}, however, these studies focussed primarily on the shift in the threshold voltage and/or properties of the device noise.

The effect of background charge polarisation in the substrate of a tunnel junction array was investigated in Ref.~\citenum{Johansson:2000} and the effect of this background was found to be non-trivial. That work involved putting a random fractional charge on a random number of islands in the array and allowing it to come to equilibrium at zero bias. However, our work concerns the transport properties at small fixed bias. 
Experiments thus far in this regime are consistent with maximal disorder behaviour \cite{Vogt:2015}. In sufficiently long devices maximal disorder is expected to completely suppress the $U$-dependence of the transport properties by self-averaging. The devices fabricated so far however are too short for full self-averaging as seen in standard condensed matter systems. It is therefore important to study the $U$-dependence in short arrays even in the case of maximal disorder.

Improvements in fabrication technologies may reduce disorder in these devices. Therefore, it is important to understand the effects of varying levels of disorder and to which degree disorder would have to be reduced to see real improvements in the behaviour of these devices. 
We investigate charge filling factors by simulating a junction array model in the charging energy limit (negligible Josephson energy) with zero disorder. We then extend this model to include various strengths of background charge disorder and consider the effect disorder has on the charge filling structure.  We specifically focus on the signatures of filling factors and background disorder in the current-voltage characteristics of the device.

\subsection*{Clean array}
\begin{figure}[th!]
\centering
\includegraphics[width=0.55\textwidth]{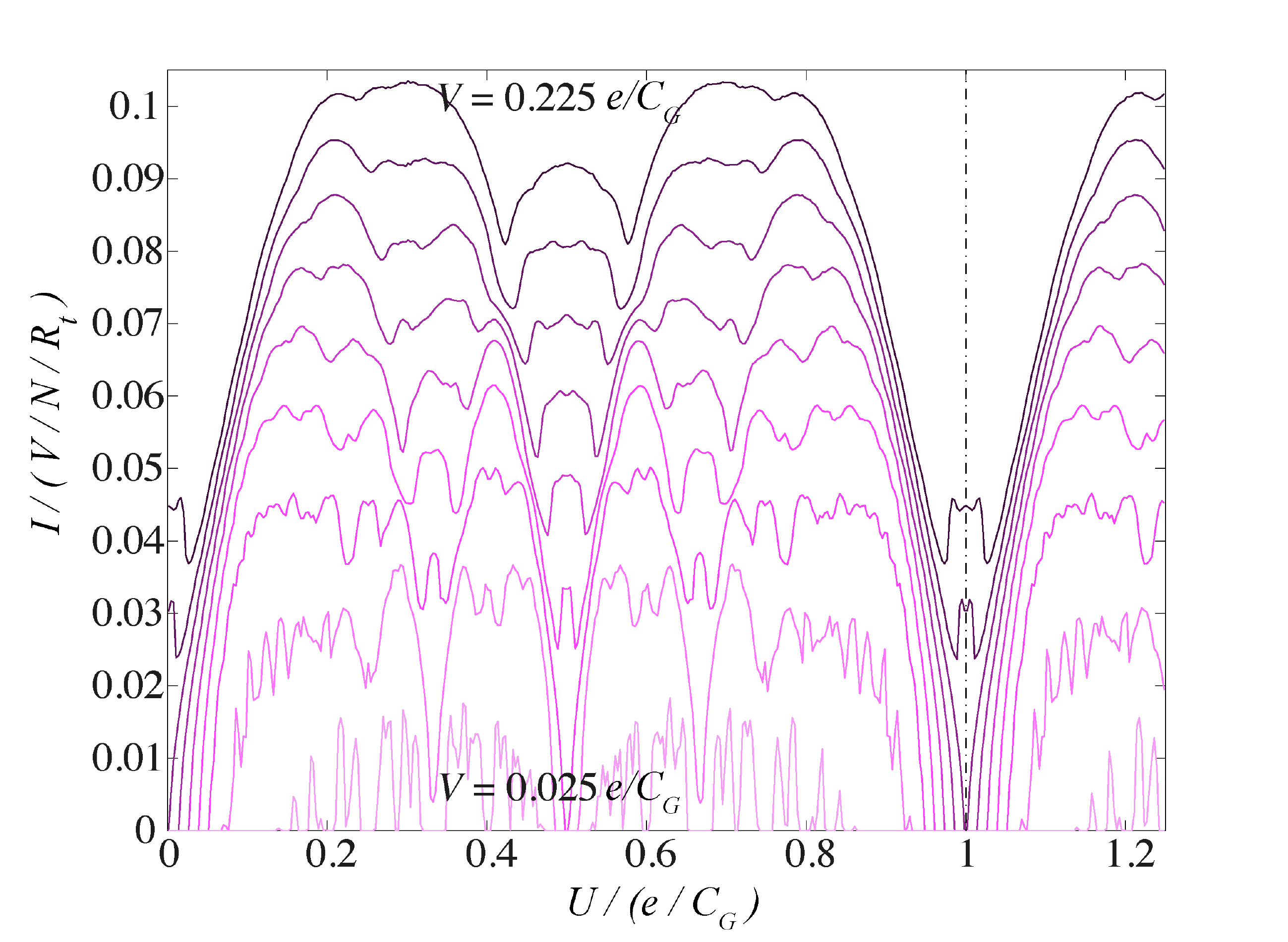}
\caption{\label{fig:Vfixed}Current response as a function of $U$ and $V$ in steps of 0.025 $e/C_G$. The fine structure is sharpest at approximately $V=0.1$ $e/C_G$. While the finer charge filling structure is lost at large $V$, this structure is preserved for smaller $V$ before becoming washed out by noise. The period of the current in $U$ is given by $e/C_G$ (dashed-dotted line).
} 
\end{figure}

We begin by considering an ideal model wherein the array is clean and homogeneous - there is no background charge disorder. We consider a one-dimensional linear array consisting of $N=50$ islands each with a junction capacitance $C_J=50$ aF, an effective capacitance to ground $C_G=2$ aF and an effective electron temperature of $T_e=30$ mK. The array is driven by a symmetric voltage bias consisting of a small fixed bias $V$ and varying offset voltage $U$, see Fig.~\ref{fig:Vfixed}. The current-voltage characteristics are non-monotonic functions of $U$ due to charge filling factors within the array. We consider the normal conducting regime where the current is comprised entirely of electrons due to single electron tunnelling. The charge interaction length can be approximated by $\Lambda=\sqrt{C_J/C_G}$, where $\Lambda=5$ (or $\Lambda=15$, where specified) for our simulations. The Hamiltonian consists of a charge interaction term and two source terms and is given by
\begin{equation}
H = \frac{1}{2}\vec{q}^T\mathbf{C}^{-1}\vec{q} + \vec{\delta_{1}}\mathbf{C}^{-1}\vec{q} C_J\bigg(\frac{V}{2}+U\bigg) + \vec{\delta_N}\mathbf{C}^{-1}\vec{q}C_J\bigg(-\frac{V}{2}+U\bigg)
\end{equation}
 for a particular charge configuration $\vec{q}$. The two source terms $\vec{\delta_{1}}$ and $\vec{\delta_{N}}$ are equal to one for site 1 and $N$, respectively and zero everywhere else. The electrostatic interactions of the array are described by the $N\times{N}$ capacitance matrix
 \begin{equation}
\mathbf{C} = \begin{pmatrix}
C_G+2C_J & -C_J & 0 &\hdots \\
-C_J & C_G+2C_J & -C_J &\ddots \\
0 & -C_J  & \ddots & \ddots \\
\vdots  &  \ddots&\ddots\\ 
\end{pmatrix} 
\end{equation}
and the charges can incoherently tunnel from island to island parameterised by a tunnelling resistance $R_t$.
We simulate the charge dynamics of the array using the kinetic Monte Carlo (KMC) method\cite{Bakhvalov:1989,Likharev:1989,Mizugaki:2005,Walker:2013,Cole:2014}. At each Monte Carlo step, all possible single electron movements are computed. The rate for each of these transitions is given by \cite{Ingold:1992}
\begin{equation}\label{eq:normalelectronrate}
\Gamma(\delta{E})=\frac{1}{e^2R_t}\int_{-\infty}^{+\infty}dE{f}(E)[1-f(E+\delta{E})]
\end{equation}
where $\delta{E}$ is the energy difference between initial and final charge configurations and $f(E)$ is the Fermi function. In general, the influence of the circuit impedance on the transition rate can be included via P(E)-theory\cite{Ingold:1992}.  However, as $C_G/C_J\ll1$, we can consider the junction array to be a low impedance environment\cite{Grabert:1991}. In this limit, $P(E)\approx\delta(E)$, resulting in the expression given in Eq.~\ref{eq:normalelectronrate} for the transition rates. These rates are used to select which transition occurs, based on a weighted probability. The total current is then computed based on the net movement, left and right, of charges over many Monte Carlo steps. To ensure the array has equilibrated, Monte Carlo steps of $10^5$-$10^6$ are used before collecting statistics with an additional $10^5$-$10^6$ steps and ensemble averaging over multiple data sets when considering disorder. For a detailed review of the KMC algorithm, we refer the reader to Ref. \citenum{Voter:2007} or Ref. \citenum{Chatterjee:2007}. 

We can also calculate the analytical threshold voltage for a current to flow through the array (at $T_e=0$)\cite{Haviland:1996wz,Walker:2013}. In the limit of long interaction length ($\Lambda\gg1$) and assuming an initially empty array
\begin{equation}\label{eq:vthres}
V_{\rm{th}}=\frac{e}{2\sqrt{C_JC_G}}
\end{equation}
where $e$ is the elementary charge. The threshold voltage and the current response are highly variable, depending on the characteristic value of the offset voltage $U$ which sets the filling fractions within the array, as shown in Fig.~\ref{fig:3d_IVU_withdots}. This behaviour is periodic in $U$, which is the focus of this paper, and is shown for half a period via the magenta line in Fig.~\ref{fig:3d_IVU_withdots}.

\begin{figure}[th!]
\centering
\includegraphics[width=0.99\textwidth]{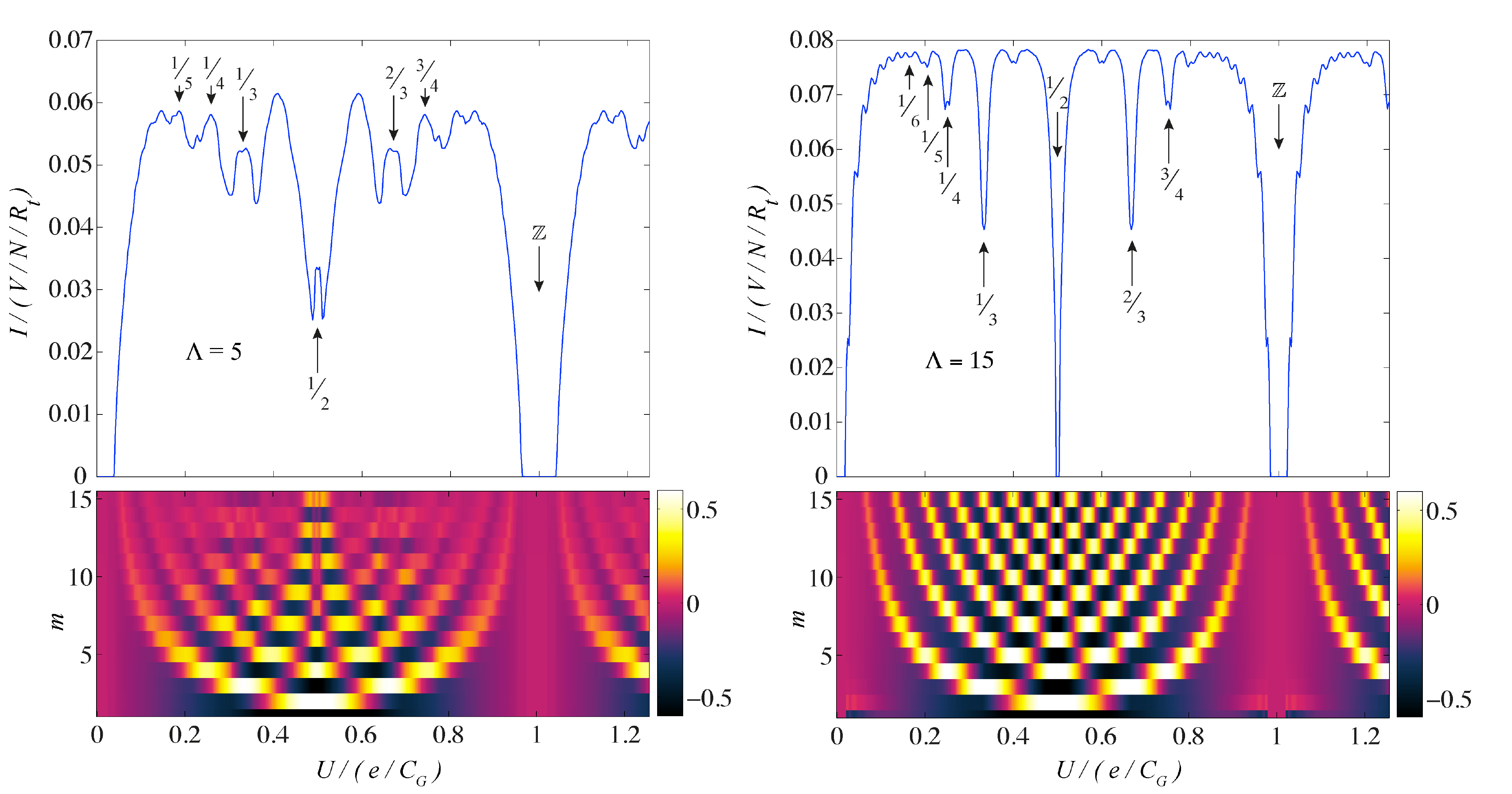}
\caption{\label{fig:Filling_factors}$I$-$U$ response at $\Lambda=5$ and $\Lambda=15$ for $V=0.1$ $e/C_G$ and $V=0.0226$ $e/C_G$, respectively. Dominant $p/q$ filling fractions are labelled for specific features. At $\Lambda=5$, mixing of different filling fractions is seen due to the short interaction length-scale. The charge-charge autocorrelation functions $\langle{n_i}(0)n_{i+m}(0)\rangle$, shown for the first 15 sites, provide a quantitative approach to identifying the filling fractions within the array. The autocorrelation shows the washing out of the filling factors as they approach $1/\Lambda=1/5$. The colour bar denotes the magnitude of the autocorrelation function.
} 
\end{figure}

The $U$-dependence can also be seen in Fig.~\ref{fig:Vfixed} where we plot the current response as a function of $U$ and $V$. The current is a periodic function of the offset voltage $U$ for all simulated values of $V$, where the period of the current in $U\propto{e/C_G}$ (the effective voltage on the ground capacitor). Therefore, the offset voltage required for an integer filled array (point of zero current) is $U_{\rm{Q-filled}}=Qe/C_G$, where $Q\in\mathbb{Z}$. Due to particle-hole symmetry, the response is symmetric about the point $U=e/2C_G$. The fine structure, seen at smaller voltages, is lost as $V$ is increased and the response flattens out and becomes more rounded. It is this fine structure that is a signature of charge filling fractions.

In order to identify filling factors ($p/q$), we compute the time averaged charge-charge autocorrelation function $\langle{n_i}(0)n_{i+m}(0)\rangle$ to measure the statistical correlations within the charge distribution. As the length of the array is not infinite, boundary effects are significant and the periodicity of the filling factors is strongest at the centre of the array.

From the current modulation in $U$, we can identify charge correlation patterns, i.e., peaks, troughs and dislocations in the patterns, as shown in Fig.~\ref{fig:Filling_factors}.  We can understand this structure by considering the filling factors of the array.  Current blockade occurs at integer filling, i.e., the array is uniformly filled and it is difficult to move this pattern forward by injecting another charge or otherwise disrupting the pattern.  In other words, integer filling factors correspond to states in which the charge configuration is rigid, therefore, the system requires significantly more energy to break this pattern and allow conduction to commence.

Fractional filling, on the other hand, allows current to flow more easily due to the presence of unoccupied sites between current carrying charges.  For a perfectly periodic charge distribution that matches the boundary conditions of the array, we also see quenching.  However, when the periodicity of the charges is disrupted, either via mismatch with the boundary or (near) degeneracy of two different charge configurations, current can flow more easily via interconversion of these charge states. As an example, periodic $p/q$ filling fractions (e.g., 1/3 filling) are more difficult to move forward, resulting in current quenching. The combination of two different filling fractions (e.g., 1/3 and 1/2 filling) is easier to transfer through the array because the two factors can be interchanged. Therefore, as a function of $U$, we see current maxima and minima as aperiodic and periodic charge patterns, respectively.  In Fig.~\ref{fig:Filling_factors}, this effect can clearly be seen for the case of $\Lambda=15$.
\begin{figure}[th!]
\centering
\includegraphics[width=0.495\textwidth]{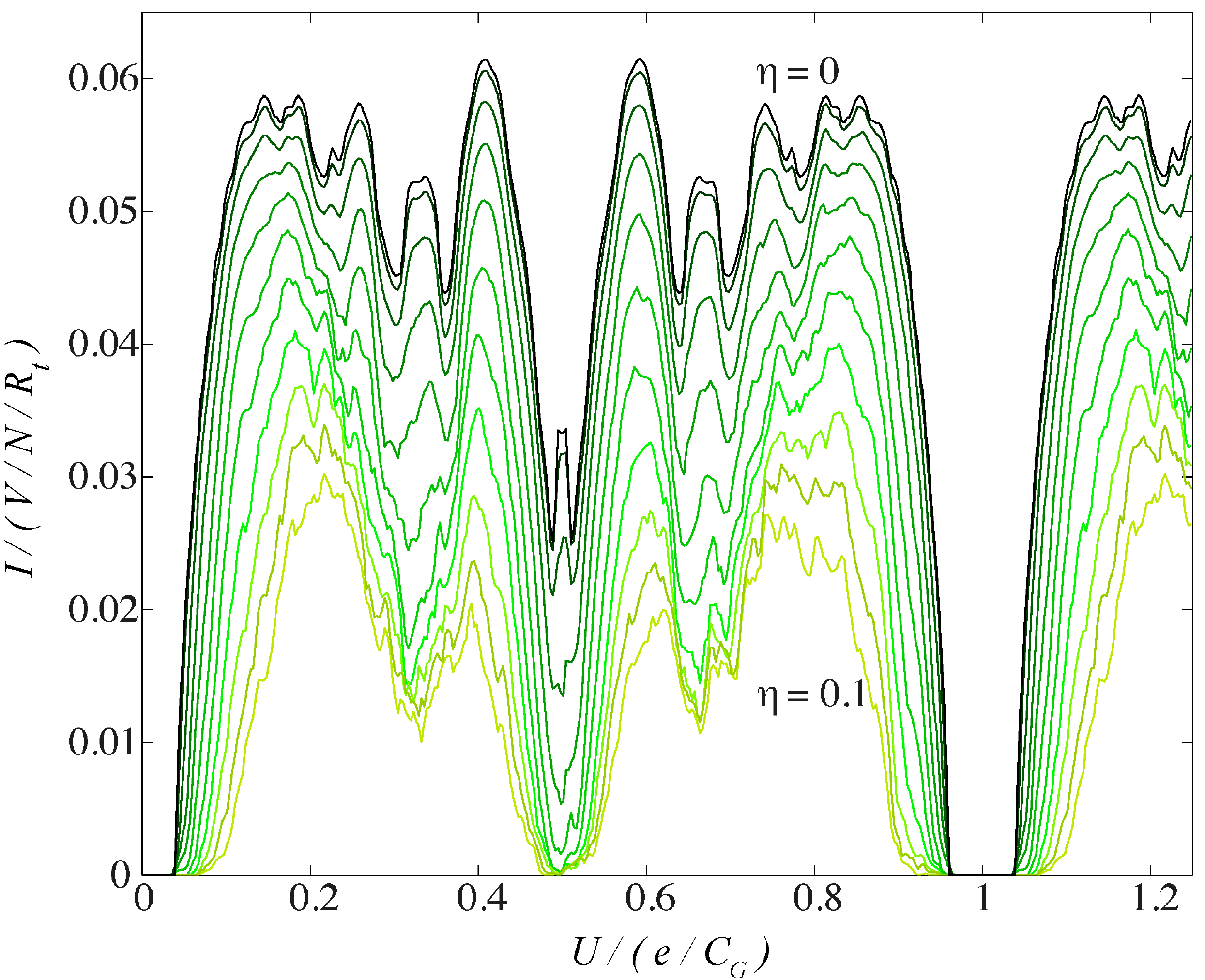}
\caption{\label{fig:disorder}$I$-$U$ characteristics at $V=0.1$ $e/C_G$ as a function of site disorder strength in steps of $0.01$ from $\eta=0$ to $\eta=0.1$. Each plot is ensemble averaged over 50 different disorder realisations. Even at very low disorder ($\eta=0.01$), while the structure remains largely unchanged, the current magnitude is reduced. The magnitude of the current continues to decrease with increasing disorder strength.
} 
\end{figure}

However, when the average spacing between charges approaches the interaction length $\Lambda$, we see a greater mixing of the filling fractions and the troughs in the current magnitude shift. As $p/q\rightarrow1/\Lambda$, we see peaks rather than troughs at rational values of $p/q$. This can be more clearly seen in Fig.~\ref{fig:Filling_factors} for the case of $\Lambda=5$.  When the mean charge separation is $<1/\Lambda$, for example, 1/2, mixing of the charge configurations is weaker. Mixing at small $\Lambda=5$ can also be seen in the autocorrelation functions in Fig.~\ref{fig:Filling_factors}, where the fractions are no longer clear and become blurred, in contrast with the clarity of the same filling factors in the autocorrelation for large $\Lambda=15$.

We observe splittings associated with defects in the dominant patterns, for instance, at $\Lambda=5$, $1/2$ filling in Fig.~\ref{fig:Filling_factors}. 
The way in which these patterns match against the boundary conditions splits the equivalency of the different degenerate configurations. For example, at $\Lambda=5$, 1/2 filling, we see that slightly above or below $p/q=1/2$ gives a slightly lower current (the trough-peak-trough structure). This feature is not seen at $\Lambda=15$, $p/q=1/2$ because at large interaction lengths averaging over a greater number of sites washes out this effect.

This is controlled by how easy it is to break the pattern, i.e., by adding an additional electron (or lack thereof) to make the pattern shift by half a period.  As an example, the (metastable) state $\vec{q}^T=10101011010101$ is slightly harder to move along the array than the `perfect' pattern of $\vec{q}^T=10101010101010$. Therefore, depending on the matching of the boundary conditions and the presence of defects changes whether we see a peak or a trough. 
\subsection*{Effects of disorder}
Experimentally, imperfections or a degree of disorder is always prevalent in the system (e.g., in the substrate and within the junctions themselves), therefore, the inclusion of disorder makes our analysis more applicable to experimental observations. Furthermore, an important question is which features of this analysis are preserved in an inhomogeneous array. In this section, we extend our model by including various background charge disorder and investigating the weak and maximal disorder limits. We simulate boxed disorder (a uniform distribution) by adding a random fractional charge to each site, $q\in[-\eta{e},+\eta{e}]$. The maximum/minimum value of disorder on any given site is given by $\pm\eta{e}$, where $\eta$ indicates the width of the disorder distribution. This type of potential disorder simulates static background charges and is often used to study mesoscopic systems\cite{Melsen:1997,Johansson:2000,Cole:2015}.

The effect of the background on the current-voltage response can be seen even at nominal disorder (e.g., $\eta=0.01$). While the structure remains largely unchanged, the current magnitude steadily decreases, as shown in Fig.~\ref{fig:disorder}. As the magnitude of static offset charges increases, the response becomes noisier and eventually only the most prominent structure remains (e.g., integer and 1/2 integer filling). However, even at very small current magnitude (strong disorder) periodic structure in $U$ remains. The current scales as $\eta$ because disorder constricts the conduction channel, making it more difficult for charge to flow (as opposed to a clean array). At strong disorder, a larger $V$ is required to overcome the random static background charges and force current through the array.

\begin{figure}[th!]
\centering
\includegraphics[width=0.65\textwidth]{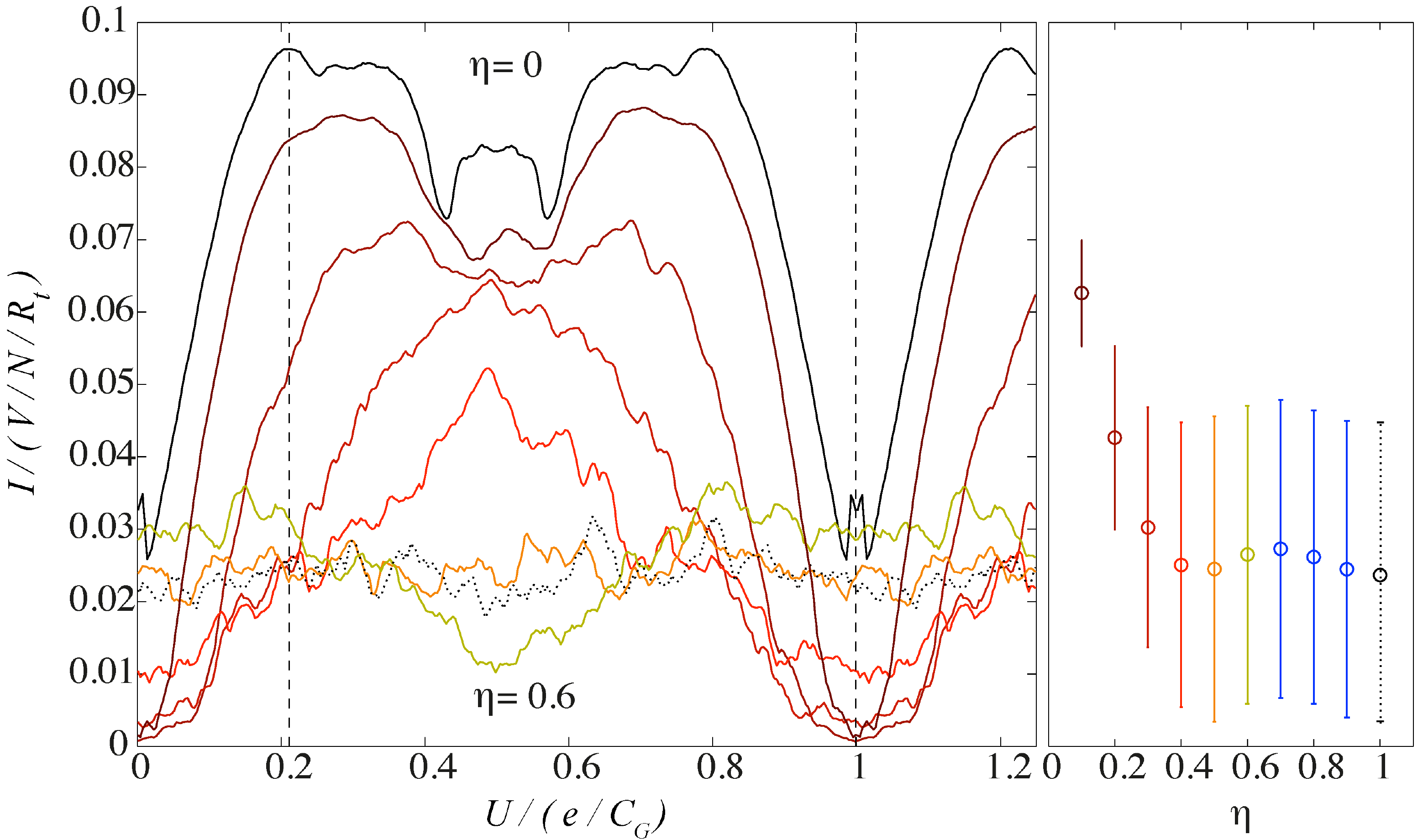}
\caption{\label{fig:disorder_plot}
$I$-$U$ response at $V=0.2$ $e/C_G$ for various disorder strengths in steps of $0.1$ from $\eta=0$ to $\eta=0.6$, $\eta=1$ is also shown (dotted line). Each plot is ensemble averaged over 50 different disorder realisations. Points of maximal and minimal current at zero disorder are shown (dashed lines). The system reaches maximal disorder at $\eta=0.5$. When $0.5<\eta<1$, the current exhibits oscillatory behaviour, however, the system reaches maximal disorder again at $\eta=1$. The mean current across $U$ and the variance in the disorder realisations for $0.1\leq\eta\leq1$ is shown via error bars which depict one standard deviation.
} 
\end{figure}

\begin{figure}[th!]
\centering
\includegraphics[width=0.6\textwidth]{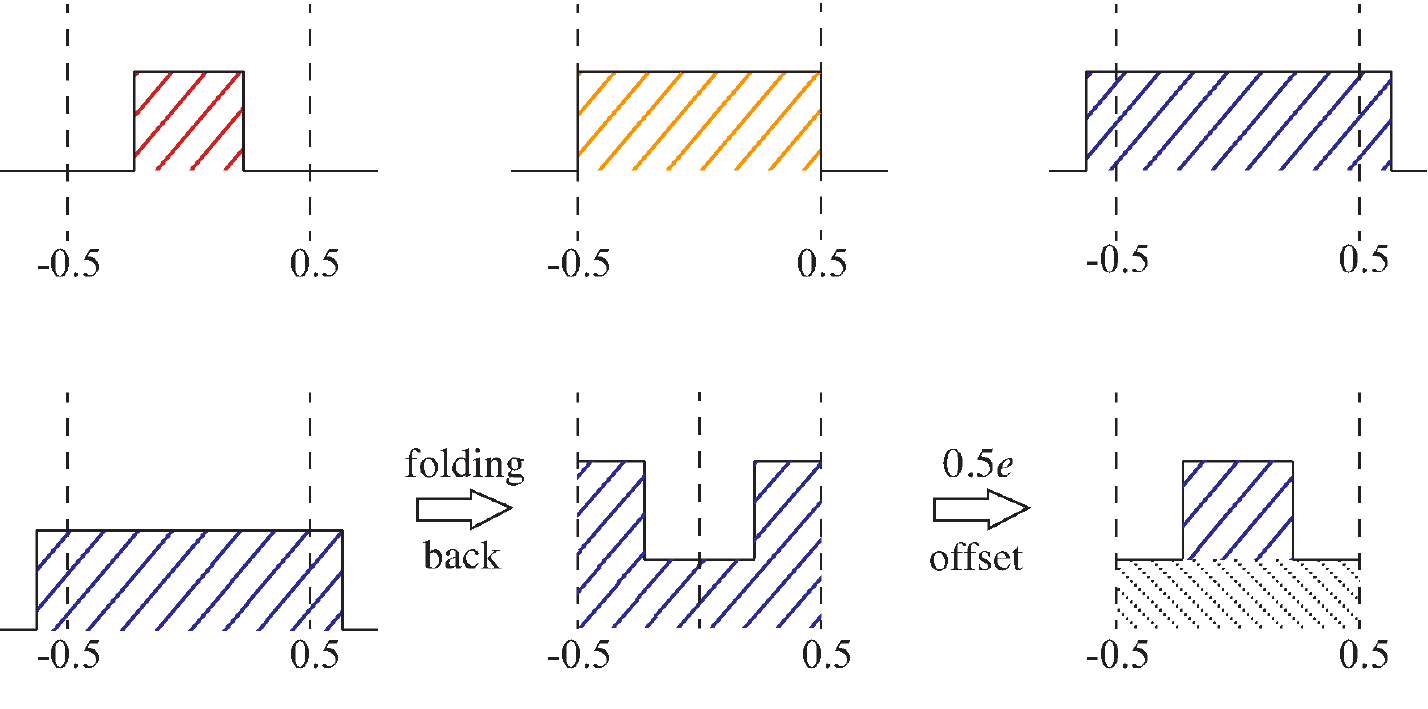}
\caption{\label{fig:nicos_diagrams}Pictorial representation of the probability distribution of the disorder as a function of $\eta$ showing maximal disorder and the folding of the charge disorder back in on itself to maximal disorder ($\eta=0.5$).  For very broad distributions, $\eta\gg0.5$, the $\eta$-dependence will abate due to decreasing `base' to `top' ratio.
}
\end{figure} 
Maximal disorder should be reached at $\eta=0.5$ as this value of disorder corresponds to the point where any value larger than 0.5 can be negated by one additional charge tunnelling onto the island. Indeed, at $\eta\approx0.5$, our model shows maximal disorder and all of the structure is lost and charge filling factors are no longer seen. In Fig.~\ref{fig:disorder_plot}, we show that the current is approximately constant across $U$ at $\eta=0.5$, indicating that maximal disorder has been reached (for this value of $V$) and a lack of $U$-dependence. Then at $\eta=0.6-0.9$, the current oscillates again, with opposite curvature (on average) to the $\eta\le0.4$ currents, with the current once again becoming approximately constant at $\eta=1$. In order to understand this behaviour, we consider three different values of disorder; less than, equal to and greater than maximal disorder, depicted in Fig.~\ref{fig:nicos_diagrams}. When the disorder distribution is broader than maximal disorder ($\eta=0.5$), adding an individual whole charge causes the distribution to wrap back around (or fold back) into the interval $\pm0.5$. This folded disorder distribution can be mapped back to the original box-distribution form with an additional `base' contribution spanning the whole $\left[-0.5, 0.5\right]$ interval by a shift of $0.5 e$ (corresponding to the opposite current curvature observed in Fig.~\ref{fig:disorder_plot}).
\begin{figure}[th!]
\centering
\includegraphics[width=0.54\textwidth]{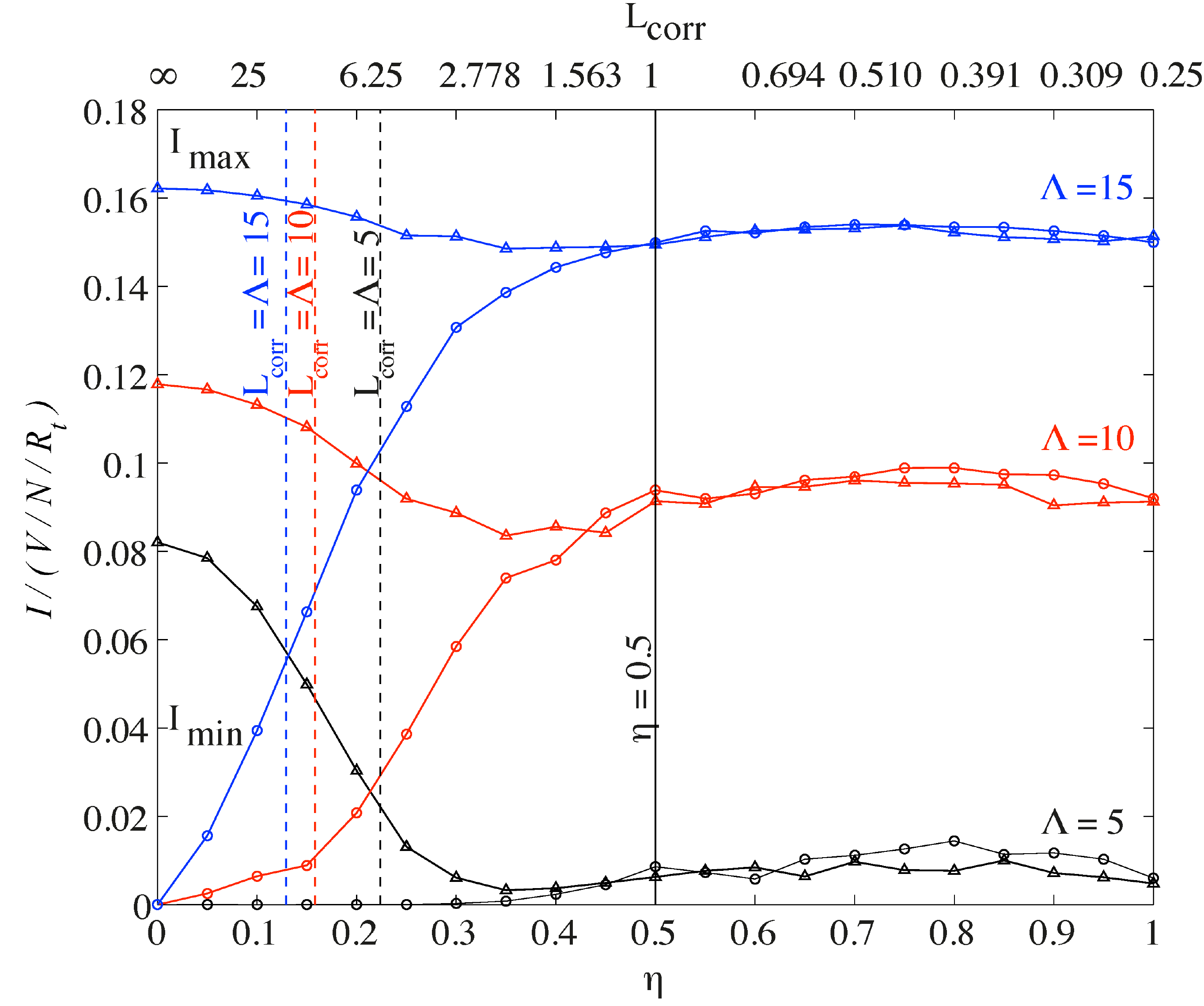}
\caption{\label{fig:I-eta}Current response as a function of $\eta$ for $\Lambda=5,10,15$, ensemble averaged over 100 different disorder realisations. The applied voltage is scaled such that $V=0.8V_{\mathrm{th}}$ for each value of $\Lambda$. The current response is shown for both $I_{\rm{max}}$ (triangles) and $I_{\rm{min}}$ (circles), i.e., the points of maximal and minimal (zero) current, respectively, in the $I$-$U$ response. The values of $L_{\rm{corr}}$ corresponding to the three values of $\Lambda$ (dashed lines) and maximal disorder (solid line) are also shown.
}
\end{figure}

In Fig.~\ref{fig:I-eta}, we plot the current response as a function of $\eta$ for three different values of $\Lambda$ at both the maximal and minimal $I$-$U$ points. The applied voltage is scaled such that $V=0.8{V_{\rm{th}}}$ for each value of $\Lambda$, where $V_{\rm{th}}$ is calculated via Eq.~\ref{eq:vthres}. In the clean array limit, the maximum current depends on $\Lambda$ because the current response in the transport regime is not exactly Ohmic, but shows a linear current-voltage dependence with a non-zero current offset. The maximum/minimum currents as a function of $U$ for all studied values of $\Lambda$ equilibrate at approximately $\eta=0.5$, i.e., maximal disorder. This is because disorder suppresses the current variations as a function of $U$, seen in Fig.~\ref{fig:disorder_plot}. The oscillatory behaviour of the current at $0.5<\eta<1$ can also be seen, with the currents at $\eta=0.5$ and $\eta=1$ equal for each value of $\Lambda$ due to maximal disorder folding. Note that throughout this discussion $\Lambda<N$.

To better understand the role of the background charge disorder, we introduce the concept of a disorder correlation length, the distance over which charges on adjacent sites are still correlated.  Beyond this length, the background disorder has the effect of randomising the electrostatic environment that the charges see and therefore, islands separated by more than this length can be considered independent. Assuming that the cumulative effect of charge disorder can be thought of as a random walk style process\cite{Fukuyama:1978}, we define the disorder correlation length as
\begin{equation}\label{eq:lcorr}
\eta\sqrt{L_{\rm{corr}}}=\frac12
\end{equation}
where $\eta$ is the value of charge disorder. When $\eta\geq0.5$, $L_{\rm{corr}}\le1$, which corresponds to a delta-correlated disorder since the minimal length-scale of the system is the inter-site-distance $L=1$. Such a delta-correlated disorder corresponds to the maximal disorder model\cite{Fukuyama:1978,Brazovskii:2004,Vogt:2015}. Conversely, $\eta\rightarrow0$ corresponds to $L_{\rm{corr}}\rightarrow\infty$, i.e., the clean array limit.

When $V$ is constant, $L_{\rm{corr}}=20,35,50$ have near identical $I$-$U$ characteristics as a function of $N$ (data not shown). In contrast to the $N$-dependence of the threshold voltage\cite{Vogt:2015a}, the system exhibits similar behaviour whether $L_{\rm{corr}}$ is equal to, less or greater than $N$. At $N=L_{\rm{corr}}$, the onset of transport changes from a boundary to a bulk effect and consequently, the dependence of $V_{\rm{th}}$ on $N$ changes. Here, however, we are considering a tunnelling array that is already in the transport regime and the modulation of the current due to a change in filling factors is naturally a bulk effect.

\section*{Discussion}
In conclusion, we have studied charge filling factors as they manifest in a (unrealistic) clean tunnel junction array. These effects are predicted to be difficult to observe experimentally as the intrinsic disorder in experimental devices likely approaches or exceeds the limit for their observation. However, while small fractional filling factors are expected to be unobservable in present day devices, our model predicts that observation and reproducibility of larger factors may be possible (excluding maximally disordered arrays). For example, current blockade near integer filling should be quite robust to static offset charges. Signatures of charge filling factors could also be observable in other one-dimensional systems. The clean array analysis is applicable to arrays of quantum phase slip elements~\cite{Mooij:2015} where the effects of localised flux disorder should be much weaker and should display longer disorder correlation lengths. The maximal disorder analysis is more applicable to Josephson junction arrays, where background charge offsets dominate the behaviour.

\section*{Acknowledgements}

We thank T. Duty and K. Cedergren for useful comments and discussions. This work was supported by the Victorian Partnership for Advanced Computing (VPAC). 

\section*{Author contributions statement}
K.W., N.V. and J.C. all contributed to this work. All authors reviewed and commented on the manuscript. 

\section*{Additional information}

Competing financial interests: The authors declare no competing financial interests.


\begin{thebibliography}{10}
\expandafter\ifx\csname url\endcsname\relax
  \def\url#1{\texttt{#1}}\fi
\expandafter\ifx\csname urlprefix\endcsname\relax\def\urlprefix{URL }\fi
\providecommand{\bibinfo}[2]{#2}
\providecommand{\eprint}[2][]{\url{#2}}

\bibitem{BenJacob:1985}
\bibinfo{author}{Ben-Jacob, E.} \& \bibinfo{author}{Gefen, Y.}
\newblock \bibinfo{title}{{New quantum oscillations in current driven small
  junctions}}.
\newblock \emph{\bibinfo{journal}{Phys. Lett. A}}
  \textbf{\bibinfo{volume}{108}}, \bibinfo{pages}{289--292}
  (\bibinfo{year}{1985}).

\bibitem{Averin:1986}
\bibinfo{author}{Averin, D.~V.} \& \bibinfo{author}{Likharev, K.~K.}
\newblock \bibinfo{title}{{Coulomb blockade of single-electron tunneling, and
  coherent oscillations in small tunnel junctions}}.
\newblock \emph{\bibinfo{journal}{J. Low Temp. Phys.}}
  \textbf{\bibinfo{volume}{62}}, \bibinfo{pages}{345--373}
  (\bibinfo{year}{1986}).

\bibitem{Likharev:1989}
\bibinfo{author}{Likharev, K.~K.}, \bibinfo{author}{Bakhvalov, N.~S.},
  \bibinfo{author}{Kazacha, G.~S.} \& \bibinfo{author}{Serdyukova, S.~I.}
\newblock \bibinfo{title}{{Single-electron tunnel junction array: an
  electrostatic analog of the Josephson transmission line}}.
\newblock \emph{\bibinfo{journal}{IEEE Trans. Magn.}}
  \textbf{\bibinfo{volume}{25}}, \bibinfo{pages}{1436--1439}
  (\bibinfo{year}{1989}).

\bibitem{Delsing:1989ea}
\bibinfo{author}{Delsing, P.}, \bibinfo{author}{Likharev, K.~K.},
  \bibinfo{author}{Kuzmin, L.~S.} \& \bibinfo{author}{Claeson, T.}
\newblock \bibinfo{title}{{Time-correlated single-electron tunneling in
  one-dimensional arrays of ultrasmall tunnel junctions}}.
\newblock \emph{\bibinfo{journal}{Phys. Rev. Lett.}}
  \textbf{\bibinfo{volume}{63}}, \bibinfo{pages}{1861--1864}
  (\bibinfo{year}{1989}).

\bibitem{Delsing:1990ik}
\bibinfo{author}{Delsing, P.}, \bibinfo{author}{Claeson, T.},
  \bibinfo{author}{Likharev, K.~K.} \& \bibinfo{author}{Kuzmin, L.~S.}
\newblock \bibinfo{title}{{Observation of single-electron-tunneling
  oscillations}}.
\newblock \emph{\bibinfo{journal}{Phys. Rev. B}} \textbf{\bibinfo{volume}{42}},
  \bibinfo{pages}{7439--7449} (\bibinfo{year}{1990}).

\bibitem{Starmark:1999}
\bibinfo{author}{Starmark, B.}, \bibinfo{author}{Henning, T.},
  \bibinfo{author}{Claeson, T.}, \bibinfo{author}{Delsing, P.} \&
  \bibinfo{author}{Korotkov, A.~N.}
\newblock \bibinfo{title}{{Gain dependence of the noise in the single electron
  transistor}}.
\newblock \emph{\bibinfo{journal}{J. Appl. Phys.}}
  \textbf{\bibinfo{volume}{86}}, \bibinfo{pages}{2132--2136}
  (\bibinfo{year}{1999}).

\bibitem{Krupenin:2000}
\bibinfo{author}{Krupenin, V.~A.}, \bibinfo{author}{Presnov, D.~E.},
  \bibinfo{author}{Zorin, A.~B.} \& \bibinfo{author}{Niemeyer, J.}
\newblock \bibinfo{title}{{Aluminum single electron transistors with islands
  isolated from the substrate}}.
\newblock \emph{\bibinfo{journal}{J. Low Temp. Phys.}}
  \textbf{\bibinfo{volume}{118}}, \bibinfo{pages}{287--296}
  (\bibinfo{year}{2000}).

\bibitem{Middleton:1993}
\bibinfo{author}{Middleton, A.~A.} \& \bibinfo{author}{Wingreen, N.~S.}
\newblock \bibinfo{title}{{Collective Transport in Arrays of Small Metallic
  Dots}}.
\newblock \emph{\bibinfo{journal}{Phys. Rev. Lett.}}
  \textbf{\bibinfo{volume}{71}}, \bibinfo{pages}{3198--3201}
  (\bibinfo{year}{1993}).

\bibitem{Melsen:1997}
\bibinfo{author}{Melsen, J.~A.}, \bibinfo{author}{Hanke, U.},
  \bibinfo{author}{M{\"u}ller, H.~O.} \& \bibinfo{author}{Chao, K.~A.}
\newblock \bibinfo{title}{{Coulomb blockade threshold in inhomogeneous
  one-dimensional arrays of tunnel junctions}}.
\newblock \emph{\bibinfo{journal}{Phys. Rev. B}} \textbf{\bibinfo{volume}{55}},
  \bibinfo{pages}{10638--10642} (\bibinfo{year}{1997}).

\bibitem{Matsuoka:1998}
\bibinfo{author}{Matsuoka, K.~A.} \& \bibinfo{author}{Likharev, K.~K.}
\newblock \bibinfo{title}{{Shot noise of single-electron tunneling in
  one-dimensional arrays}}.
\newblock \emph{\bibinfo{journal}{Phys. Rev. B}} \textbf{\bibinfo{volume}{57}},
  \bibinfo{pages}{15613--15622} (\bibinfo{year}{1998}).

\bibitem{Mueller:1998}
\bibinfo{author}{M{\"u}ller, H.~O.}, \bibinfo{author}{Katayama, K.} \&
  \bibinfo{author}{Mizuta, H.}
\newblock \bibinfo{title}{{Effects of disorder on the blockade voltage of
  two-dimensional quantum dot arrays}}.
\newblock \emph{\bibinfo{journal}{J. Appl. Phys.}}
  \textbf{\bibinfo{volume}{84}}, \bibinfo{pages}{5603--5609}
  (\bibinfo{year}{1998}).

\bibitem{Johansson:2000}
\bibinfo{author}{Johansson, J.~R.} \& \bibinfo{author}{Haviland, D.~B.}
\newblock \bibinfo{title}{{Random background charges and Coulomb blockade in
  one-dimensional tunnel junction arrays}}.
\newblock \emph{\bibinfo{journal}{Phys. Rev. B}} \textbf{\bibinfo{volume}{63}},
  \bibinfo{pages}{014201} (\bibinfo{year}{2000}).

\bibitem{Vogt:2015}
\bibinfo{author}{Vogt, N.} \emph{et~al.}
\newblock \bibinfo{title}{{One-dimensional Josephson junction arrays: Lifting
  the Coulomb blockade by depinning}}.
\newblock \emph{\bibinfo{journal}{Phys. Rev. B}} \textbf{\bibinfo{volume}{92}},
  \bibinfo{pages}{045435} (\bibinfo{year}{2015}).

\bibitem{Bakhvalov:1989}
\bibinfo{author}{Bakhvalov, N.~S.}, \bibinfo{author}{Kazacha, G.~S.},
  \bibinfo{author}{Likharev, K.~K.} \& \bibinfo{author}{Serdyukova, S.~I.}
\newblock \bibinfo{title}{{Single-electron solitons in one-dimensional tunnel
  structures}}.
\newblock \emph{\bibinfo{journal}{Sov. Phys. JETP}}
  \textbf{\bibinfo{volume}{95}}, \bibinfo{pages}{1010--1021}
  (\bibinfo{year}{1989}).

\bibitem{Mizugaki:2005}
\bibinfo{author}{Mizugaki, Y.} \& \bibinfo{author}{Shimada, H.}
\newblock \bibinfo{title}{{Monte Carlo study of charge transport in slantingly
  coupled arrays of small tunnel junctions}}.
\newblock \emph{\bibinfo{journal}{Phys. Rev. B}} \textbf{\bibinfo{volume}{71}},
  \bibinfo{pages}{115103} (\bibinfo{year}{2005}).

\bibitem{Walker:2013}
\bibinfo{author}{Walker, K.~A.} \& \bibinfo{author}{Cole, J.~H.}
\newblock \bibinfo{title}{{Correlated charge transport in bilinear tunnel
  junction arrays}}.
\newblock \emph{\bibinfo{journal}{Phys. Rev. B}} \textbf{\bibinfo{volume}{88}},
  \bibinfo{pages}{245101} (\bibinfo{year}{2013}).

\bibitem{Cole:2014}
\bibinfo{author}{Cole, J.~H.}, \bibinfo{author}{Lepp{\"a}kangas, J.} \&
  \bibinfo{author}{Marthaler, M.}
\newblock \bibinfo{title}{{Correlated transport through junction arrays in the
  small Josephson energy limit: incoherent Cooper-pairs and hot electrons}}.
\newblock \emph{\bibinfo{journal}{New J. Phys.}} \textbf{\bibinfo{volume}{16}},
  \bibinfo{pages}{063019} (\bibinfo{year}{2014}).

\bibitem{Ingold:1992}
\bibinfo{author}{Ingold, G.-L.} \& \bibinfo{author}{Nazarov, Y.~V.}
\newblock \bibinfo{title}{{Charge Tunneling Rates in Ultrasmall Junctions}}.
\newblock In \bibinfo{editor}{Grabert, H.} \& \bibinfo{editor}{Devoret, M.~H.}
  (eds.) \emph{\bibinfo{booktitle}{Single Charge Tunneling: Coulomb Blockade
  Phenomena in Nanostructures}}, \bibinfo{pages}{21--107}
  (\bibinfo{publisher}{Plenum}, \bibinfo{address}{New York},
  \bibinfo{year}{1992}).

\bibitem{Grabert:1991}
\bibinfo{author}{Grabert, H.} \bibinfo{author}{\emph{et.al.}}
\newblock \bibinfo{title}{{Single electron tunneling rates in multijunction circuits}}.
\newblock \emph{\bibinfo{journal}{Z. Phys. B: Condens. Matter}} \textbf{\bibinfo{volume}{84}},
  \bibinfo{pages}{143--155} (\bibinfo{year}{1991}).
  
\bibitem{Voter:2007}
\bibinfo{author}{Voter, A.~F.}
\newblock \emph{\bibinfo{title}{{Introduction to the Kinetic Monte Carlo
  Method}}}, vol. \bibinfo{volume}{235} of \emph{\bibinfo{series}{Nato Science
  Series, Series II: Mathematics, Physics And Chemistry}}
  (\bibinfo{publisher}{Radiation Effects in Solids}, \bibinfo{year}{2007}).

\bibitem{Chatterjee:2007}
\bibinfo{author}{Chatterjee, A.} \& \bibinfo{author}{Vlachos, D.~G.}
\newblock \bibinfo{title}{{An overview of spatial microscopic and accelerated
  kinetic Monte Carlo methods}}.
\newblock \emph{\bibinfo{journal}{J. Comput.-Aided Mater. Des.}}
  \textbf{\bibinfo{volume}{14}}, \bibinfo{pages}{253--308}
  (\bibinfo{year}{2007}).

\bibitem{Haviland:1996wz}
\bibinfo{author}{Haviland, D.~B.} \& \bibinfo{author}{Delsing, P.}
\newblock \bibinfo{title}{{Cooper-pair charge solitons: The electrodynamics of
  localized charge in a superconductor}}.
\newblock \emph{\bibinfo{journal}{Phys. Rev. B}} \textbf{\bibinfo{volume}{54}},
  \bibinfo{pages}{6857--6860} (\bibinfo{year}{1996}).

\bibitem{Cole:2015}
\bibinfo{author}{Cole, J.~H.}, \bibinfo{author}{Heimes, A.},
  \bibinfo{author}{Duty, T.} \& \bibinfo{author}{Marthaler, M.}
\newblock \bibinfo{title}{{Parity effect in Josephson junction arrays}}.
\newblock \emph{\bibinfo{journal}{Phys. Rev. B}} \textbf{\bibinfo{volume}{91}},
  \bibinfo{pages}{184505} (\bibinfo{year}{2015}).

\bibitem{Fukuyama:1978}
\bibinfo{author}{Fukuyama, H.} \& \bibinfo{author}{Lee, P.~A.}
\newblock \bibinfo{title}{{Dynamics of the charge-density wave. I. Impurity
  pinning in a single chain}}.
\newblock \emph{\bibinfo{journal}{Phys. Rev. B}} \textbf{\bibinfo{volume}{17}},
  \bibinfo{pages}{535--541} (\bibinfo{year}{1978}).

\bibitem{Brazovskii:2004}
\bibinfo{author}{Brazovskii, S.} \& \bibinfo{author}{Nattermann, T.}
\newblock \bibinfo{title}{{Pinning and sliding of driven elastic systems: from
  domain walls to charge density waves}}.
\newblock \emph{\bibinfo{journal}{Adv. Phys.}} \textbf{\bibinfo{volume}{53}},
  \bibinfo{pages}{177--252} (\bibinfo{year}{2004}).

\bibitem{Vogt:2015a}
\bibinfo{author}{Vogt, N.}, \bibinfo{author}{Cole, J.~H.} \&
  \bibinfo{author}{Shnirman, A.}
\newblock \bibinfo{title}{{Depinning of disordered bosonic chains}}.
\newblock \emph{\bibinfo{journal}{arXiv: 1510.01383v1}} (\bibinfo{year}{2015}).

\bibitem{Mooij:2015}
\bibinfo{author}{Mooij, J.~E.} \emph{et~al.}
\newblock \bibinfo{title}{{Superconductor-insulator transition in nanowires and
  nanowire arrays}}.
\newblock \emph{\bibinfo{journal}{New J. Phys.}} \textbf{\bibinfo{volume}{17}},
  \bibinfo{pages}{033006} (\bibinfo{year}{2015}).
\end{thebibliography}
\end{document}